\documentclass[aps,epsfig,floatfix,twocolumn,amssymb,superscriptaddress,amsmath,notitlepage]{revtex4-1} 

\usepackage{amsmath,amsthm,latexsym,amssymb,amsfonts,mathtools,epsfig}
\usepackage{graphicx,texdraw,natbib,subfigure,subeqnarray,fancyhdr,amsmath,multirow,color,bm,mathrsfs}
\usepackage[english]{babel}
\usepackage[utf8]{inputenc}			

\newcommand{\pd}[2]{\frac{\partial #1}{\partial #2}} 
\newcommand{\td}[2]{\frac{\mathrm{d} #1}{\mathrm{d} #2}} 

\newcommand{\de}{\mathrm{d}}

\newcommand{\br}{\mathbf{r}}

\newcommand{\bs}[1]{\boldsymbol{#1}}
\newcommand{\avg}[1]{\left\langle{#1}\right\rangle}

\usepackage[normalem]{ulem}
\usepackage{cancel}

\begin{document}

\title{Autophoretic motion in three dimensions} 

\author{Maciej Lisicki}
\email{m.lisicki@damtp.cam.ac.uk}
\affiliation{Department of Applied Mathematics and Theoretical Physics, University of Cambridge, Wilberforce Road, Cambridge UK.}
\affiliation{Institute of Theoretical Physics, Faculty of Physics, University of Warsaw, Poland.}

\author{Shang Yik Reigh}
\affiliation{Max Planck Institute for Intelligent Systems, Stuttgart, Germany.}

\author{Eric Lauga}
\email{e.lauga@damtp.cam.ac.uk}
\affiliation{Department of Applied Mathematics and Theoretical Physics, University of Cambridge, Wilberforce Road, Cambridge UK.}
\date{\today}

\begin{abstract}

Janus particles with the ability to move phoretically in self-generated chemical concentration gradients are model systems for active matter. Their motion typically consists of straight paths with rotational diffusion being the dominant reorientation mechanism. 
In this paper, we show theoretically that by a suitable surface coverage of both activity and mobility,  translational and rotational motion can be induced arbitrarily in three dimensions. The resulting trajectories are in general helical, and their pitch and radius can be controlled by adjusting the angle between the translational and angular velocity. Building on the classical mathematical framework for axisymmetric self-phoretic motion under fixed-flux chemical boundary condition, we first show how to calculate the most general three-dimensional  motion for an arbitrary surface coverage of a spherical particle. After illustrating our results on  surface distributions,  we next introduce a simple intuitive patch model to serve as a guide for designing arbitrary phoretic spheres. 
 
 \end{abstract} 
\maketitle

\section{Introduction}

Living systems of swimming microorganisms exhibit a rich variety of collective phenomena \cite{Koch2011}, including swarming motion \cite{Turner2010} and collective oscillations on scale much larger than the individual cell length \cite{Chen2017}. Suspensions of bacteria may exhibit non-linear rheological properties \cite{Lopez2015} or enhanced diffusion \cite{Leptos2009}. The properties of these complex biological suspensions can be studied effectively in model artificial systems. The ability to manufacture and control biomimetic systems tailored to specific applications has become one of the key challenges of modern nanotechnology \cite{Katuri2017}. 

Catalytic particles are now well established as model system to address the properties of living and active matter. To a certain degree they can also be controlled by external  fields \cite{Mano2017} to mimic bacterial run-and-tumble motion. Catalytic micro-motors can also be used for microscopic cargo transport \cite{Baraban2012} and exhibit collective dynamics which shares many features with living systems, such as clustering \cite{Theurkauff2012}, swarming and structure formation due to imposed anisotropic interactions \cite{Yan2016,Dileonardo2016}.  In order to  describe these collective effects, however, we have to understand the basic building block of such a system, namely single-particle dynamics.

The mechanism of propulsion of individual catalytic particles relies on the phoretic motion in self-generated gradients  \cite{anderson1984}. These may involve electric field (electrophoresis), temperature field (thermophoresis) or chemical concentration (diffusiophoresis) \cite{anderson1989,Moran2016}.  While the underlying formalism holds for all these types of motion, we focus our attention here on the latter case. The concentration gradients are typically produced by covering the surface of the body by a layer of catalyst \cite{Lattuada2011}. The non-uniform concentration field along the surface drives a diffusive flow, leading to an effective slip-flow on the surface of the particle \cite{Golestanian2005}. In result, motion of the particle itself is induced, with the leading-order flow field of a dipolar character due to the absence of external forces and torques, which is akin to the swimming character of living microorganisms \cite{Lauga2009}. The effectiveness of the catalyst is quantified by the chemical activity of its surface. The resulting surface slip flow is proportional to the local concentration gradient via a mobility coefficient, which is related to the details of the local particle-solute interaction potential \cite{anderson1989}. These two material properties - activity and mobility - fully characterise the flow generated on the surface.

The ability to produce the gradients and resulting self-phoretic motion requires a certain level of asymmetry in the system. There are two basic ways to achieve this: (i) an asymmetry in surface properties of a spherical particle, or (ii) an asymmetry of the shape of the particle.

The classical generation of motion by patterning can be achieved chemically in the so-called Janus particles by coating a cap of the spherical particle by a layer of catalyst. Following pioneering work on bi-metallic rods \cite{Paxton2004},  early theoretical approaches   explored the relations between the cap coverage by catalyst and the resulting propulsion \cite{Golestanian2005,Golestanian2007}. In this case, activity and mobility patterns were axially symmetric, and thus able to generate motion along the symmetry axis of the particle. This idea has inspired a number of experimental realisations, including designing platinum-coated spherical colloids in hydrogen peroxide solution \cite{Howse2007,Ebbens2011,Ebbens2012} (although the detailed physical mechanism is still debated \cite{Brown2014,Reigh2016}) and vapour-induced motion of gold-iridium spheres \cite{Dong2015}. Theoretical efficiency of this propulsion mechanism has also been studied theoretically  \cite{Sabass2010,Sabass2012}.

The role of geometry in inducing the gradients can be exploited by non-spherical shapes of the particles, such as analytically treatable oblate or prolate spheroids \cite{michelin2017,Popescu2010},dimers of spheres uniformly covered by activity and mobility \cite{Michelin2015,Reigh2015}, or asymmetric L-shaped particles\cite{Kummel2013}.  The same effect can be used to produce flow confined by chemically active asymmetric boundaries, enabling local small-scale pumping of the fluid \cite{michelin2015tom,lisicki2016}. In these cases, the concentration and thus flow fields have in general to be determined numerically \cite{Rueckner2007,Montenegro-Johnson2015}.

All axially symmetric designs proposed in the aforementioned studies produce unidirectional motion, with possible reorientations of the particle purely due to rotational Brownian motion or external forcing. The need for designing rotational phoretic swimmers led to the idea of Janus particles conjoined in dimers capable of propelling on helical paths \cite{Majee2017}. Uneven surface properties and shape imperfections due to manufacturing have also been shown experimentally to lead to translational and rotational motion of Janus spheres at an air-water interface\cite{Wang2017}.  The same effect, however, can be achieved with a perfectly spherical particle, provided that an asymmetric coating pattern is used.  In a recent study,  glancing angle metal evaporation was used onto a colloidal crystal to break the symmetry of the catalytic patch due to shadowing by neighbouring colloids  \cite{Archer2015}. This allowed to produce batches of phoretic swimmers with a well defined rotational speed. The particles have been characterised in terms of the resulting rotation. As a next step, in this work we provide the formal description of the relation between the coverage and resulting motion that allows to predict the properties of such patterned objects.

Inspired by these experimental advances and ideas, we develop here the
mathematical formalism capable of predicting the translational and rotational
velocity of a spherical phoretic particle with a given surface activity and mobility
coverage. Following the classical framework \cite{Golestanian2005,Golestanian2007}, we model the surface activity by imposing a local chemical flux boundary
condition on the surface of the particle. We assume that the diffusion of solute molecules is fast compared to
advection and reaction rates and thus  consider the limit of vanishing P\'{e}clet and Damk\"ohler number for the solute, $\mathrm{Pe}=\mathrm{Da}=0$.  
By expanding the activity and mobility in spherical harmonics, we calculate the resulting surface slip flow which drives the motion and the swimming kinematics. We next  introduce a conceptually simpler patch model, in which the motion is induced by pairs of interacting patches of activity and mobility. Due to the bilinear mathematical nature of the flow generation problem and the associate boundary conditions, the interaction of patches can be superposed to predict the motion resulting from a given collection of point sources and patches of activity. Considering finite-sized domains instead does not change the qualitative picture but only modifies the quantitative characteristics of motion. We demonstrate the basic ingredients needed to program the particles to move along straight lines, circles, and arbitrary helical trajectories.

The paper is organised as follows. In Sec.~\ref{1model} we describe the underlying model for the purely diffusive dynamics of solute transport around the particle arising from its surface activity, and  relate it to the effective slip flow caused by the presence of concentration gradients. For an arbitrary surface pattern of activity and mobility, we determine in Sec.~\ref{2analytics} the general formulae for the resulting velocity and angular velocity of the sphere by averaging the local slip flow over the particle surface. In Sec.~\ref{3programmed}, we apply these results to analyse a few model coverage patterns resulting in a prescribed type of motion, including pure rotation or combination of translation and rotation. In order to explore in detail the minimal requirements for the generation of a desired trajectory, we introduce the patch model in Sec.~\ref{4patch} and identify the effect of interactions between small patches of activity and mobility on the motion of the sphere leading in general to arbitrary helical trajectories. We conclude the paper in Sec.~\ref{6conclusions}. In 
Appendix~\ref{5kinematic}, we derive the classical result of exact trajectories for given translational and angular velocity.


\section{The model}\label{1model}

\begin{figure}[ht]
 \centering
\includegraphics[width=0.35\textwidth]{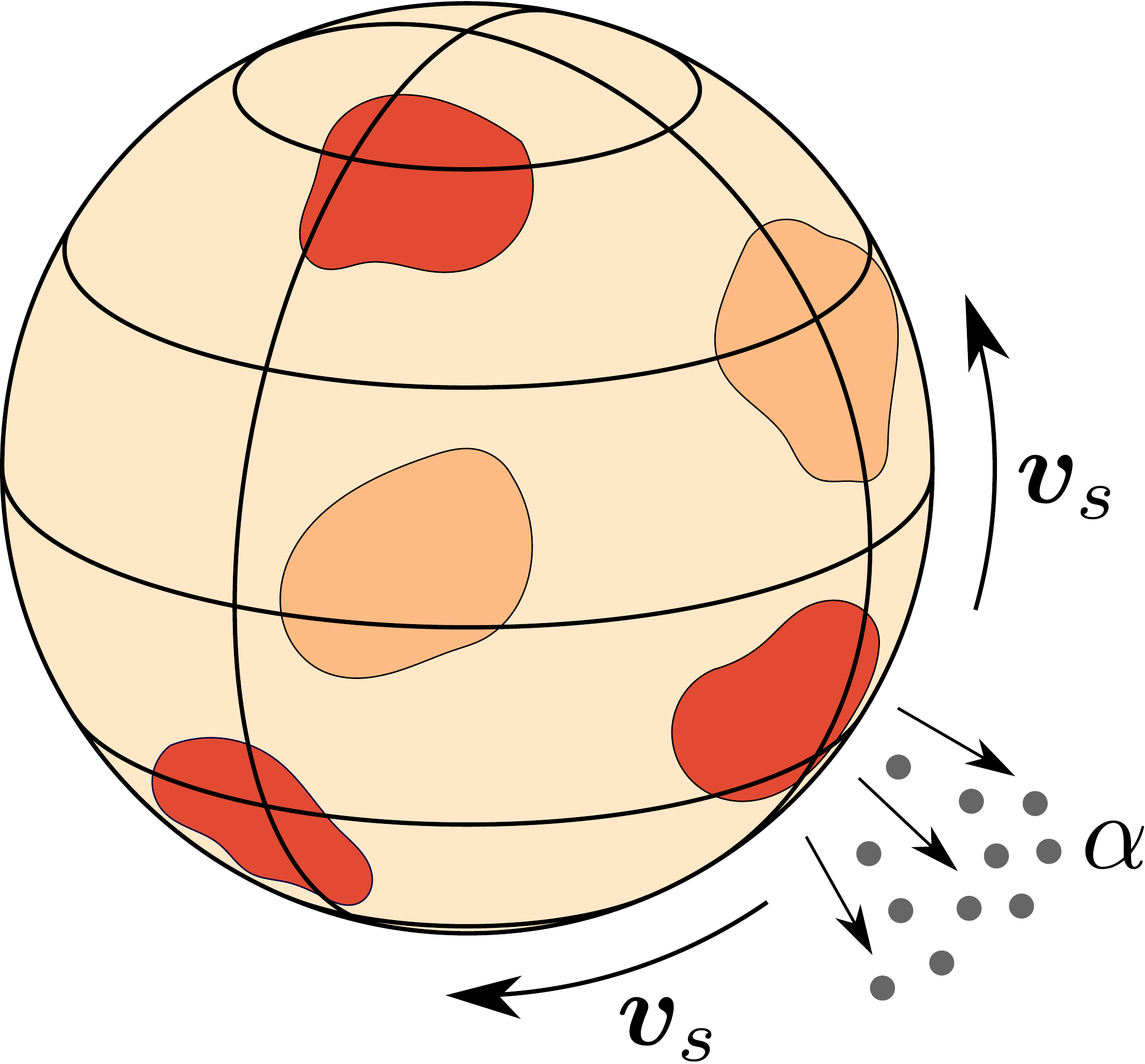}
\caption{Sketch of a spherical microparticle with a non-uniform surface distribution of chemical activity and mobility, which is here represented by patches of inreased/decreased activity (dark red) and mobility (light orange). Due to the chemical reaction, gradients of concentration of solute particles $\alpha$ are produced, which induce osmotic flows along the surface. This in turn leads to phoretic translational and rotational motion of the sphere.
}
\label{system}
\end{figure}

A spherical particle of radius $a$ is immersed in a Newtonian solvent of density $\rho$
and dynamic viscosity $\eta$ (see Fig. \ref{system}). 
We consider only one kind of chemical species $\alpha$ in solutions for
simplicity and the particle interacts with these molecules by imposing a fixed chemical activity $\mathcal{A}$ along its surface.
The chemical activity is inhomogeneous on the particle surface and hence
  concentration gradients of solute particles are created,  inducing  surface slip flows and globally the motion of the particle with a characteristic velocity $V$.
The fluid flow in the system is characterised by the Reynolds number
$\mathrm{Re} = \rho V a/\eta$.

The chemical profile around the particle is generally affected by both advection and diffusion of solute. The relative importance of the two is quantified by the P\'{e}clet number, $\mathrm{Pe}= Va/D$, with $D$ being the diffusivity of molecules $\alpha$ in the solvent. In typical diffusiophoretic experiments involving colloids \cite{Howse2007,Ebbens2011,Paxton2004}, the small sizes of particles ($a\sim 1$ $\mu\mathrm{m}$) and the very fast diffusion of   solute molecules  render the P\'{e}clet number negligible, and thus the dynamics of the concentration field can be described as purely diffusive with a very good approximation.

The catalytic surface is assumed to prescribe a non-uniform surface activity denoted 
$\mathcal{A}(\theta,\phi)$ using   spherical polar coordinates $(r,
\theta,\phi)$ in the body-fixed frame.
We consider the far-field concentration of $\alpha$ molecules to be $c_0$. 

The chemical reactions are modelled by a fixed-flux boundary
condition at the particle surface~\cite{Moran2016,Michelin2014}, where the
solute molecules are set to be constantly absorbed or emitted at the particle surface depending on the choice of
magnitudes of activity as $\mathcal{A}>0$ and $\mathcal{A}<0$, respectively.
Introducing the relative concentration $c=c_\alpha - c_0$,   the
steady state concentration field in the low P\'{e}clet number regime is the solution of Laplace's
equation,
\begin{equation}\label{laplace}
\nabla^2 c = 0,
\end{equation}
subject to the flux boundary condition on the particle surface 
\begin{equation}\label{flux}
-\bs{J}\cdot \bs{e}_r \big|_{r=a} = \mathcal{A}(\theta,\phi),
\end{equation}
where $\bs{e}_r$ is the unit normal (radial) vector on the surface of the sphere,
and a constant concentration at infinity, i.e. $c(r\to\infty) = 0$. The diffusive flux is given by the Fick's law and reads
\begin{equation}
\bs{J}=-D\nabla c.
\end{equation}
Hence, a positive relative concentration indicates emission of solute molecules at the surface of the particle  while negative values correspond to  absorption.

The flow around a particle is generated by  gradients in concentration around the surface. Specifically, the diffusive flow $\bs{v}_s$ tangential to the surface is proportional to the local tangential gradients of concentration as
\begin{equation} \label{slipvelocity}
\bs{v}_s =  \mathcal{M}(\theta,\phi) (\bs{1}-\bs{e}_r\bs{e}_r)\cdot\nabla c,
\end{equation}
where $\mathcal{M}(\theta,\phi)$ is called the surface mobility, which depends  on the details of particle-solvent interaction forces \cite{anderson1989}. Given the typical propulsion velocities of $V\sim 10$ $\mu\mathrm{m}/\mathrm{s}$ \cite{Paxton2004}, the Reynolds number  $\mathrm{Re}\sim 10^{-5}$ is negligible, and the flow in the reference frame of the particle can be found by solving the incompressible Stokes equations
\begin{equation}
\eta \nabla^2 \bs{v} = \nabla p, \qquad \nabla\cdot\bs{v} = 0,
\end{equation}
with the boundary condition at infinity being the self-propulsion velocity of the particle,  
\begin{equation}
\bs{v}(\bs{r}\to\infty) \sim - (\bs{V} + \bs{\Omega}\times\bs{r}).
\end{equation}

We aim at determining the resulting velocity, $\bs{V}$, and angular velocity, $\bs{\Omega}$, given the arbitrary coverage fields $\mathcal{A}(\theta,\phi)$ and $\mathcal{M}(\theta,\phi)$, thus  extending the generic propulsion mechanism proposed in 
Ref.~\cite{Golestanian2007} to three-dimensional motion. The propulsion velocity and angular velocity can be obtained from the reciprocal theorem by averaging the slip flow over the surface of the sphere as \cite{anderson1984,anderson1989,Stone1996}
\begin{equation} \label{V_avg}
\bs{V} = - \langle \bs{v}_s \rangle,
\end{equation}
and
\begin{equation} \label{O_avg}
\bs{\Omega} = \frac{3}{2a} \langle \bs{v}_s\times\bs{e}_r \rangle,
\end{equation}
where the surface average  reads explicitly
\begin{equation}
\langle ... \rangle = \frac{1}{4\pi a^2} \oint (...)\ \mathrm{d}S
\end{equation}
with the spherical surface element $\de S = a^2 \sin\theta\de\theta \de \phi = -a^2 \de \zeta \de\phi$, where $\zeta=\cos\theta$.

In summary, the prescribed coverage fields $\mathcal{A}(\theta,\phi)$ and
$\mathcal{M}(\theta,\phi)$ determine the resulting three-dimensional
translational and rotational motion of the particle. After solving  
Laplace's equation for the concentration field, we use it as input to calculate the slip flow, which is finally averaged over the  surface of the particle. We note that this is possible due to the decoupling of flow and concentration fields in the limit of low P\'{e}clet numbers. If this is not the case, the solute is additionally advected by the flow \cite{Michelin2014} and both problems are fully coupled.

\section{Analytical results}\label{2analytics}

To analyse the effect of coverage geometry, we expand both the activity and mobility surface patterns in spherical harmonics series and derive a general expression for the translational and rotational velocity in terms of the expansion coefficients. 

The general solution of Eq.~\eqref{laplace} can be classically written in terms of spherical harmonics in the body-fixed frame as
\begin{equation}\label{concentration}
c = - \sum_{\ell=0}^{\infty}\sum_{m=-\ell}^{\ell} C_{\ell m}{r^{-\ell-1}} Y_{\ell m}(\theta,\phi),
\end{equation}
with the polar angle {$\theta\in[0,\pi]$} and the azimuthal angle $\phi\in[0,2\pi)$, where the spherical harmonic of order $\ell$ and azimuthal number $m$ is defined as 
\begin{equation} 
Y_{\ell m}(\theta,\phi) = N_{\ell m} P_\ell^m (\zeta) e^{im\phi},
\end{equation}
with $\zeta=\cos\theta$ while $P_\ell^m$ are the associated Legendre polynomials,
and we have used the normalisation constant
\begin{equation}
N_{\ell m} = \sqrt{{(2\ell+1)}\frac{(\ell-m)!}{(\ell+m)!}}\cdot
\end{equation}
By expanding the activity distribution in spherical harmonics, we can write
\begin{equation}\label{activity_exp}
\mathcal{A}(\theta,\phi) = \sum_{\ell m} A_{\ell m} Y_{\ell m}(\theta,\phi),
\end{equation}
where we have introduced a shorthand notation for the double sum in
Eq.~\eqref{concentration} and which we will use consistently
hereafter. Using the boundary condition in Eq.~\eqref{flux}, we find the relation between the expansion coefficients
\begin{equation}
C_{\ell m} = - \frac{a^{\ell+2}}{D(\ell+1)}A_{\ell m}.
\end{equation}
We expand the mobility in an analogous manner
\begin{equation}
\mathcal{M}(\theta,\phi)= \sum_{\ell m} M_{\ell m} Y_{\ell m}(\theta,\phi),
\end{equation}
and the sets of coefficients $A_{\ell m}$ and $M_{\ell m}$ are prescribed for a given coverage pattern of the particle. 

The surface velocity may now be expressed using 
Eq.~\eqref{slipvelocity} together with the expansions of activity $\mathcal{A}$ and mobility $\mathcal{M}$. For the characteristics of motion to be evaluated, we need to find the surface slip flow, which is proportional to the mobility times the tangential gradient of concentration on the surface at $r=a$. From Eq.~\eqref{slipvelocity}, it is given by
\begin{align}\label{eq:16}
\bs{v}_s=& \frac{1}{D} \sum_{\ell m}\sum_{\ell' m'} \frac{A_{\ell m}M_{\ell' m'}}{\ell+1}Y_{\ell' m'}  \left(\pd{ Y_{\ell m}}{\theta}\bs{e}_\theta + \frac{1}{\sin\theta}\pd{ Y_{\ell m}}{\phi}\bs{e}_\phi \right)  .
\end{align}

The expression in Eq.~\eqref{eq:16} needs to be directly averaged over the angles to yield the
translational velocity (with a minus sign) or averaged after taking a cross
product with $\bs{e}_r$ to obtain the rotational velocity. It is most
convenient to express the velocity components in the Cartesian body-fixed
frame $(x,y,z)$. 
Using the relationships between unit vectors in spherical
  polar coordinates,
$\bs{e}_r\times\bs{e}_\theta = \bs{e}_\phi$ and $\bs{e}_r \times\bs{e}_\phi =
-\bs{e}_\theta$, we find the local angular velocity increment
\begin{align}
\bs{v}_s\times\bs{e}_r =& \frac{1}{D} \sum_{\ell m}\sum_{\ell' m'} \frac{A_{\ell m}M_{\ell' m'}}{\ell+1}Y_{\ell' m'} \left( \frac{1}{\sin\theta}\pd{ Y_{\ell m}}{\phi}\bs{e}_\theta -\pd{ Y_{\ell m}}{\theta}\bs{e}_\phi \right).
\end{align}

Representing the basis vectors in Cartesian coordinates, we can evaluate the averages with respect to the azimuthal angle $\phi$ by noting that spherical harmonics are orthonormal in the azimuthal numbers $p,q$, since
\begin{equation}
\avg{e^{i(p-q)\phi}}_\phi := \frac{1}{2\pi}\int_{0}^{2\pi} e^{i(p-q)\phi}\de\phi = \delta_{pq}.
\end{equation}
This allows to eliminate the summation over $m'$ in favour of $m$. The averages with respect to the polar angle $\theta$ cannot be easily evaluated but can be explicitly expressed as averages of combinations of associate Legendre polynomials.

The results below are valid for an arbitrary coverage, described by the spherical harmonics expansion coefficients of activity $\mathcal{A}$ ($A_{\ell m}$) and mobility $\mathcal{M}$ ($M_{\ell m}$). Denoting  $\alpha_{\ell m} = A_{\ell m} N_{\ell m}$ and $\mu_{\ell m} = M_{\ell m} N_{\ell m}$ (no summation convention), the translational velocity is finally found by taking the real part (indicated by $\Re$)
\begin{widetext}
\begin{align} \label{vx}
V_x &={\Re}\left\{\frac{1}{2D}\sum_{\ell m \ell'}\frac{\alpha_{\ell m}}{\ell+1}\Big[K_{-}(\ell m, \ell' m_1)\mu_{\ell' m_1} + K_{+}(\ell m, \ell' m_2)\mu_{\ell' m_2}  \Big]\right\}, \\ \label{vy}
V_y &={\Re}\left\{\frac{i}{2D}\sum_{\ell m \ell'}\frac{\alpha_{\ell m}}{\ell+1}\Big[K_{-}(\ell m, \ell' m_1)\mu_{\ell' m_1} - K_{+}(\ell m, \ell' m_2)\mu_{\ell' m_2}  \Big]\right\}, \\ \label{vz}
V_z &={\Re}\left\{\frac{1}{D}\sum_{\ell m}\frac{(-1)^{m+1}\alpha_{\ell+1,m}}{2\ell+3}\Bigg[\frac{\ell+m+1}{2\ell+1} \mu_{\ell,-m}  -\frac{\ell+1}{\ell+2}\frac{\ell-m+2}{2\ell+5} \mu_{\ell+2,-m}\Bigg]\right\},
\end{align}
where $m_1=-(m-1)$ and $m_2=-(m+1)$. Interestingly, in the limit
of an axisymmetric pattern for both mobility and activity, we find
$V_x=V_y=0$, and $V_z$ reduces to the classical expression   \citep{Golestanian2007}. The angular velocity follows as
\begin{align} 
 \Omega_x &=-{\Re}\left\{\frac{3i}{4a D}\sum_{\ell m\ell'}\frac{\alpha_{\ell m}}{\ell+1} \Big[L_{-}(\ell m, \ell' m_1)\mu_{\ell' m_1} - L_{+}(\ell m, \ell' m_2)\mu_{\ell' m_2}  \Big]\right\}, \label{omegax} \\ 
 \label{omegay}
\Omega_y &={\Re}\left\{\frac{3}{4a D}\sum_{\ell m\ell'}\frac{\alpha_{\ell m}}{\ell+1}\Big[L_{-}(\ell m, \ell' m_1)\mu_{\ell' m_1} + L_{+}(\ell m, \ell' m_2)\mu_{\ell' m_2}  \Big]\right\}, \\ \label{omegaz}
\Omega_z & ={\Re}\left\{ \frac{3i}{2a D}\sum_{\ell m}\frac{ m(-1)^{m+1} \alpha_{\ell,m}\mu_{\ell,-m} }{(\ell+1)(2\ell+1)}\right\}, 
\end{align}
\end{widetext}
 For an axisymmetric pattern ($m=0$), we have no rotational motion, as $\bs{\Omega}={\bf 0}$, as expected by symmetry.

The auxiliary integrals
\begin{align}
K_{\pm}(\ell m, \ell' m') = J_1(\ell m, \ell' m') \pm m J_2(\ell m, \ell' m'), \\
L_{\pm}(\ell m, \ell' m') = J_3(\ell m, \ell' m') \pm m J_4(\ell m, \ell' m'),
\end{align}
contain averages with respect to the polar angle $\theta$.  Introducing the notation
\begin{align}
\avg{\ldots}_\zeta &= \frac{1}{2}\int_{-1}^{1} \de \zeta (\ldots)
\end{align}
with $\zeta=\cos\theta$, we have
\begin{align}
J_1(\ell m, \ell' m') &= \avg{  \zeta\sqrt{1-\zeta^2} \td{P_{l}^{m}}{\zeta} P_{\ell'}^{m'} }_\zeta , \\
J_2(\ell m,\ell' m') &= \avg{ \frac{1}{\sqrt{1-\zeta^2}}  P_{l}^{m} P_{\ell'}^{m'} }_\zeta, \\
J_3(\ell m, \ell' m') &= \avg{  \sqrt{1-\zeta^2} \td{P_{l}^{m}}{\zeta} P_{\ell'}^{m'} }_\zeta , \\
J_4(\ell m,\ell' m') &= \avg{ \frac{\zeta}{\sqrt{1-\zeta^2}}  P_{l}^{m} P_{\ell'}^{m'} }_\zeta.
\end{align}
These integrals, although easy to evaluate numerically, seem impossible to express analytically in general.

\section{Programmed phoretic motion}\label{3programmed}

The complete solution for $\bs{V}$ and $\bs{\Omega}$ presented above allows for computation of the motion characteristics for an arbitrary coverage. We shall now demonstrate examples of  simple coverage patterns which result in the desired type of motion. 

There are five cases of particular interest which we will analyse:
(a) pure translation, (b) pure rotation, (c) rotation about the translation direction, $\bs{V}\parallel\bs{\Omega}$, (d) rotation perpendicular to translation, $\bs{V}\cdot\bs{\Omega}=0$, and (e) a given arbitrary angle $\psi$ between the two vectors.

\subsection{Pure translation} This case is the simplest mode of motion and can be induced by an axisymmetric coverage of the particle by mobility $\mathcal{M}(\theta)$ and activity $\mathcal{A}(\theta)$. The high symmetry eliminates the possibility of rotations. This limit corresponds to $m=0$ in Eqs.~\eqref{vx}-\eqref{omegaz} and leads to the vanishing of all components but $V_z$ which in this case takes the form found previously   \cite{Golestanian2007}.

\subsection{Pure rotation} For the  rotational motion, axial symmetry of the
coverage pattern need be broken. In order to additionally restrict the motion
 not to have a translational component, 
we impose no dependence on the polar angle $\theta$ and the $\phi \to \pi + \phi$ (i.e. 2-fold) rotational symmetry in the azimuthal angle in both activity and mobility. However, the symmetries of the two coverages have to be different (e.g.~slightly rotated). An exemplary coverage satisfying that symmetry is an activity pattern with one plane of azimuthal symmetry,
 and a mobility pattern with the same symmetry but rotated by an angle
 $\Delta$ about the $z$ axis, e.g.
 \begin{align}
\mathcal{A}(\theta,\phi) &= A \cos^2\phi, \\
\mathcal{M}(\theta,\phi) &= M \cos^2(\phi + \Delta).
\end{align} 
Here, $A$ and $M$ set the scale for the control parameters. It is important to
note that the patterns have no dependence on the angle $\theta$ (and are thus  symmetric about the plane $\theta=\pi/2$), thus ruling out the possible motion in the $z$ direction. Any other distribution sharing the same symmetries would result in a qualitatively similar motion. 

The expansion of activity and mobility in spherical harmonic yields that only $A_{00}=A/2$, $M_{00}=M/2$ and coefficients with $m=\pm 2$  are non-zero. Moreover, we find the symmetries
\begin{align} 
 A_{\ell 2} &= A_{\ell -2} = A \tilde{A}_{\ell 2} \label{symmetriesA}, \\ 
M_{\ell 2} &=M \tilde{M}_{\ell 2} e^{2i\Delta}, \nonumber \\ 
M_{\ell -2}&=M \tilde{M}_{\ell 2} e^{-2i\Delta}, \nonumber
\end{align}
where $\tilde{A}_{\ell m}$ and $\tilde{M}_{\ell m}$ are are sets of known coefficients. The detailed expressions are omitted here. Since only indices $m=-2,0,2$ have non-zero coefficients, we find from 
Eqs.~\eqref{vx}-\eqref{vy} and \eqref{omegax}-\eqref{omegay} that the $x,y$ components of $\bs{V}$ and $\bs{\Omega}$ must vanish as they connect only indices differing by $\pm1$. Since there is no dependence on the polar angle, the only  coefficients with even $\ell$ are present in the expansions of $\mathcal{A}$ and $\mathcal{M}$, which is the reason for the vanishing value of $V_z$. We are left with  evaluating explicitly $\Omega_z$ from Eq.~\eqref{omegaz} as
\begin{equation} \label{omegazspin}
\Omega_z = - \frac{MA}{Da} C\sin 2\Delta.
\end{equation}
where the constant can be evaluated using the known coefficients as $C=6\sum_{\ell} \frac{ \tilde{A}_{\ell 2} \tilde{M}_{\ell
    2}}{\ell+1}\approx 0.406$. 
Importantly, we see that when activity and mobility share exactly the same symmetry, that is when $\Delta=0$, there is no rotational motion. However, an offset causes the sphere to rotate without translating, and the rotational velocity for small angles is proportional to $\Delta$. The maximal value of the rotational speed, $MA C/Da$ is obtained when $\Delta=\pi/4$. At $\Delta=\pi/2$ the motion ceases again by symmetry. 

\subsection{Collinear rotation and translation}

A case of particular interest arises when the axis of rotation coincides with the direction of motion. To this end, we can modify the previously discussed pure rotation by disturbing the polar symmetry, i.e. distinguishing one pole of the sphere from the other. A slight modification sufficient to achieve this is given by the activity and mobility of
 \begin{align}
\mathcal{A}(\theta,\phi) &= A (1+\cos\theta)\cos^2\phi, \\
\mathcal{M}(\theta,\phi) &= M \cos^2(\phi + \Delta).
\end{align}  
The activity has the same azimuthal distribution necessary for inducing rotational motion but varies in intensity in the polar direction,  increasing towards the pole at $\theta=0$. The azimuthal numbers are still limited to $m=-2,0,2$, and thus all  $x$ and $y$ components of $\bs{V}$ and $\bs{\Omega}$ are ruled out (as Eqs \eqref{vx}-\eqref{vy} and \eqref{omegax}-\eqref{omegay} couple azimuthal numbers differing by 1). The expansions retain their symmetries as in Eqs \eqref{symmetriesA}, and thus Eq.~\eqref{omegazspin} is still a valid expression for $\Omega_z$ with exactly the same constant $C$. {This is due to the fact the additional factor $\cos\theta$ in activity only adds terms with odd values of $\ell$ into the expansion. Since $\Omega_z$ only pairs activity and mobility coefficients with same $\ell$, and the expansion for $\mathcal{M}$ contains only terms with even $\ell$, the rotational velocity remains unaffected.}

The presence of all terms (for each $\ell$) in the expansion leads to a
non-zero translational velocity along the $z$-axis. The $m=0$ terms produce a
velocity component independent of the azimuthal angle $\phi$, equal to
$-MA/12D$. 
For comparison, if the sphere surface had no dependence on the angle $\phi$,
i.e. uniform mobility $\mathcal{M}=M$ and activity
$\mathcal{A}=A(1+\cos\theta)$, its translational velocity is equal to
$-MA/3D$. 
The terms depending on the angle $\phi$ ($m \ne 0$ terms) give a velocity
component with a $\Delta$-dependent factor, equal to $-MAT\cos(2\Delta)/12D$, so that 
in total the translational velocity reads
\begin{align}
V_z &= -\frac{MA}{12 D}\left( 1 + T \cos 2\Delta \right) 
\end{align}
where 
\begin{align}
T =& 2\sum_{\ell}\frac{\tilde{A}_{\ell+1,2}}{\sqrt{2\ell+3}}\left[\sqrt{\frac{\ell^2+2\ell+3}{2\ell+1}} \tilde{M}_{\ell 2 } \right. \\ \nonumber
&\left.- \frac{\ell+1}{\ell+2}\sqrt{\frac{\ell(\ell+4)}{2\ell+5}} \tilde{M}_{\ell+2,2}\right]
 \approx -0.175.
\end{align}
It is evident that the translational motion is weakly dependent on the particular offset and mostly determined by the global coverage pattern in the polar direction. When the rotational motion is maximised at $\Delta=\pi/4$, the ratio $ \Omega_z a/V_z = 12 C \approx 4.9$. 

\subsection{Orthogonal rotation and translation}

Another special case arises when the translational and rotational velocities are perpendicular, i.e. when $\bs{V}\cdot\bs{\Omega}=0$. The resulting motion of the particle is then along a circular periodic trajectory. To give an example of such an arrangement, we consider the following coverage pattern
\begin{align}
\mathcal{A}(\theta,\phi) &= A (1+\cos\theta), \\
\mathcal{M}(\theta,\phi) &= M (1+\cos\phi).
\end{align}
By symmetry, we expect this pattern to result in translation along the $z$-axis (the mobility symmetry axis) and rotation about the $y$-axis. Indeed, the activity pattern has only two non-zero coefficients in the spherical harmonics expansion, $\alpha_{00}=A$ and $\alpha_{10}=A/\sqrt{3}$. The mobility expansion yields the non-zero coefficients to be $\mu_{00}$ and $\mu_{2n+1,\pm 1}$ with $n\geq 0$. Computation of the auxiliary integrals $K_{\pm}$ and $L_{\pm}$ compatible with these values of polar and azimuthal numbers yields $V_{x}=V_{y}=0$ and $\Omega_{x}=\Omega_{z}=0$. For the translational velocity, we find that only one pair of coefficients ($\alpha_{10}\mu_{00}$) contributes to give finally
\begin{equation}
V_{z} = -\frac{3 MA}{D}\cdot
\end{equation}
Similarly, we find that the angular velocity about the $y$-axis is only affected by the coefficients $\alpha_{10}$ and $\mu_{1\,\pm 1}$ leading to
\begin{equation}
\Omega_{z} = -\frac{3}{8aD}{\alpha_{10}\mu_{11}} = -\frac{3 MA}{16\sqrt{2}  aD}\cdot
\end{equation}
Clearly, we have $\bs{\Omega}\cdot\bs{V}=0$, so this pattern will result in circular motion of the particle with a radius $V/\Omega  = 16\sqrt{2}a\approx 22.7a$.

\subsection{General rotation and translation}

It is evident from Eqs.~\eqref{vx}-\eqref{omegaz} that in order to induce rotation and translation in the $xy$ plane, the expansions for activity and mobility need to have coefficients with the azimuthal numbers $m$ differing by $\pm 1$, since these are paired in the  expressions for $V_{x,y}$ and $\Omega_{x,y}$. Various examples of such coverage may be found but these tend to not be  clearly intuitive due to their complicated mathematical structure. We hence turn our attention to the patch model in the next section which is sufficient to show the symmetries needed for the generation of the desired type of motion.

\section{The patch model}\label{4patch}

\subsection{Activity and mobility patches}
Suppose the active site on the surface of the sphere is a small spherical cap with $\theta<\theta_A$ near the pole around $\theta=0$ and has uniform activity $A$, while the rest is inert ($A=0$ elsewhere). This can be written mathematically as   $\mathcal{A}=A\Theta(\zeta_A-\zeta)$, with $\zeta=\cos\theta$ and $\Theta$ being the Heaviside step function.  We now perform a Legendre polynomial expansion and write the activity as $\mathcal{A} = \sum_{\ell} A_\ell P_{\ell}(\zeta)$. Using the properties of the Legendre polynomials, we find $A_\ell = A [P_{\ell-1}(\zeta_A) - P_{\ell+1}(\zeta_A)]/2= A F_{\ell}$. The tangential slip velocity from Eq.~\eqref{slipvelocity} is thus given by
\begin{equation}
\bs{v}_s = -\mathcal{M}(\theta,\phi) \frac{A}{2D}\left[\sum_{\ell} \frac{F_{\ell}}{\ell+1} \td{P_{\ell}}{\zeta} \right]\sin\theta \bs{e}_\theta,
\end{equation}
while the local angular velocity increment is given by
\begin{equation}
\bs{v}_s\times\bs{e}_r =  \mathcal{M}(\theta,\phi) \frac{A}{2D}\left[\sum_{\ell} \frac{F_{\ell}}{\ell+1} \td{P_{\ell}}{\zeta} \right] \sin\theta \bs{e}_\phi,
\end{equation}
where the mobility is still arbitrary.

Suppose now that only small patches of the surface of the sphere   can sustain a slip flow along the surface. That means that there are patches of mobility on the surface, while there is no slip flow on the rest of the surface. The mobility distribution can then be described as a collection of approximate Dirac delta functions scattered on the sphere. 

\subsection{A pair of patches}

 \begin{figure}[t]
 \centering
\includegraphics[width=0.35\textwidth]{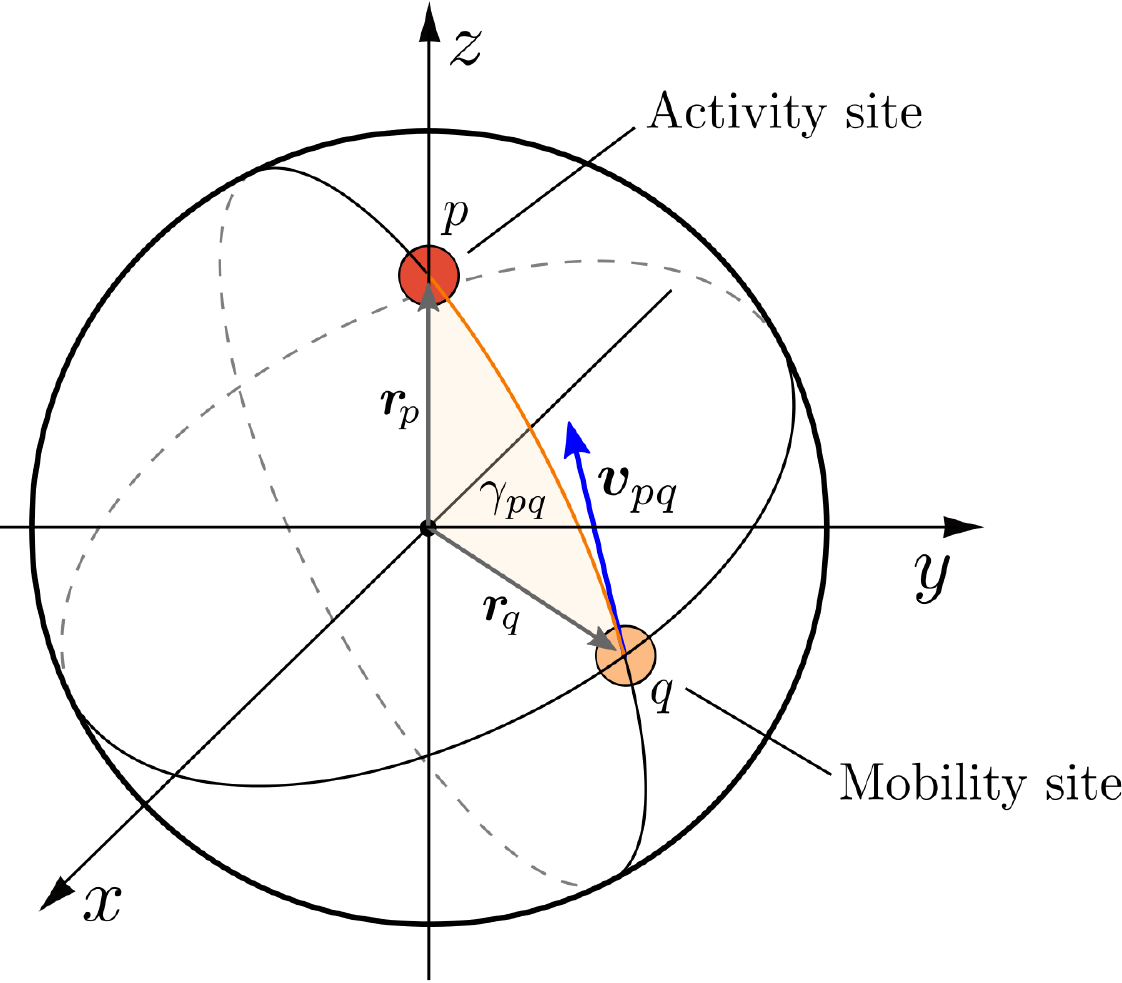}
\caption{The motion resulting from the presence of a pair of active (dark red) and mobility (light orange) sites on the surface of the particle.  The velocity increment lies in the plane spanned by the two position vectors of the patches, while the angular velocity increment resulting from this geometry is normal to this plane.}\label{pair}
\end{figure}

It is instructive to consider a pair of patches on a unit sphere: one small active site of activity $A_p$ at $\bs{r}_p$ and a singular  mobility site of mobility $M_q$ at $\bs{r}_q$, as illustrated in 
Fig.~\ref{pair}. In this case, the mobility is non-zero only at the point $q$, thus $\mathcal{M}(\theta,\phi) = M_q \delta(\bs{r}-\bs{r}_q)$.  The angle between the patches, $\gamma_{pq}$, is found by noting that  $\bs{r}_p\cdot\bs{r}_q=  \cos\gamma_{pq}$. The surface slip velocity generated by this pair, $\bs{v}_{pq}$, is tangential to the sphere at $\bs{r}_q$ and lies in the plane spanned by the vectors $\bs{r}_{p,q}$ (it is thus aligned with the local unit vector $\bs{e}_\theta (q)$), while the increment to angular velocity $\bs{v}_s\times\bs{e}_r$ is perpendicular to that plane (in the local direction of $\bs{e}_\phi(q)$). The velocities are obtained from Eqs \eqref{V_avg} and \eqref{O_avg} as
\begin{align} \label{Vpq}
\bs{V}_{pq} &= \frac{A_p M_q}{2D}  G(\gamma_{pq}) \bs{e}_\theta(q), \\ \label{Opq}
\bs{\Omega}_{pq} &= \frac{3A_p M_q}{4aD} G(\gamma_{pq}) \bs{e}_\phi(q),
\end{align}
with
\begin{equation} \label{gtheta}
G(\gamma_{pq})=\left[\sum_{\ell}\frac{F_\ell}{\ell+1}\td{P_{\ell}}{\zeta}\right]_{\gamma_{pq}}\sin\gamma_{pq},
\end{equation}
where the subscript $\gamma_{pq}$ indicates that the derivatives are evaluated
at this angle and $\zeta=\cos\theta$ refers to the polar angle in local
coordinates where $p$ is on the pole $\theta=0$.  We see from 
Eqs.~\eqref{Vpq} and \eqref{Opq} that there is no
flow generated when the patches are at opposite poles of the sphere, which is
expected by symmetry.

The increments in Eqs.~\eqref{Vpq} and \eqref{Opq} are given in local coordinate systems in which the (local) vector $\bs{e}_z(p)$ is pointing towards the active site, so they need to be transformed to the body-fixed Cartesian system to properly account for many pairs. The local unit vectors can be found using the position vectors of the mobility and activity patches as
\begin{align}
\bs{e}_\phi(q) &= \frac{\bs{r}_q\times\bs{r}_p}{|\bs{r}_q\times\bs{r}_p|}, \\
\bs{e}_\theta(q) &= \bs{e}_\phi(q) \times \br_{q}.
\end{align}

\begin{figure}[t]
 \centering
\includegraphics[width=0.5\textwidth]{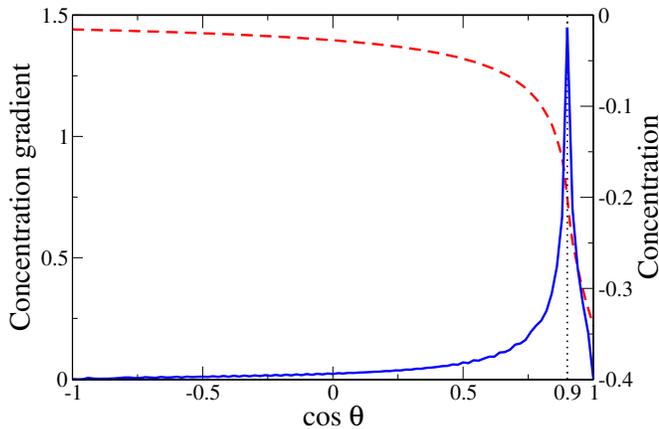}
\caption{The relative concentration field, $c=c_A-c_0$ (red dashed line), where the solute molecules $\alpha$ are absorbed at the surface ($A>0$), and
  the resulting dimensionless concentration gradient, $G(\theta)$ from Eq.~\eqref{gtheta}, (blue solid line) due to a patch of unit activity at {$\zeta>\zeta_A=0.9$}.
   The dimensionless concentration is scaled by $Aa/D$, and the gradient is scaled by $A/D$. The curves were obtained  using 300 terms in the Legendre expansion.  The rim of the active patch is marked with a dotted line. The gradient is maximal at the rim and decays rapidly towards the poles. The mobility patch positioned at a given angle picks out the value of the gradient that sets the magnitude of the velocity increment. 
}
\label{gradient}
\end{figure}

In Fig.~\ref{gradient} we present an example of  the concentration field generated by
a small activity patch with {$\zeta>\zeta_A=0.9$}, 
 along with the corresponding concentration gradient. The concentration gradient is maximal
near the rim of the patch, at which maximal propulsion and rotational
velocities are induced (see also Ref. \cite{Reigh2016}).
Increasing the angular distance between the active and mobility sites significantly decreases the concentration gradient, which is proportional to the induced slip velocity. Thus for the generation of larger linear and angular
velocities, the patches should be at a small angular distance. The resulting velocity is proportional to the product of activity and mobility of the patches.

\subsection{Linear superposition of pair contributions}
The solution of the two-patch system provided above opens the way to analyse combinations of many patches. Due to the averaging procedure which involves integrating the delta functions, many patches can be treated as a superposition of pair interactions, with the resulting velocities being sums over all activity-mobility pairs of patches $(p,q)$ 
\begin{align}
\bs{V} &= \sum_{p,q} \bs{V}_{pq}, \\
\bs{\Omega} &= \sum_{p,q} \bs{\Omega}_{pq}.
\end{align}
 We note that  linear superposition is mathematically possible because of the set of  boundary conditions. We consider Neumann boundary conditions for the concentration field and Dirichlet boundary conditions for the velocity field. With these boundary conditions, for linear superposition to be applicable  both activity and mobility patches
  have to be homogeneous (i.e~zero) outside the area where they are acting, and non-overlapping. 
 
\subsection{Patch-induced programmed motion}
  
   \begin{figure*}
 \centering
\includegraphics[width=0.8\textwidth]{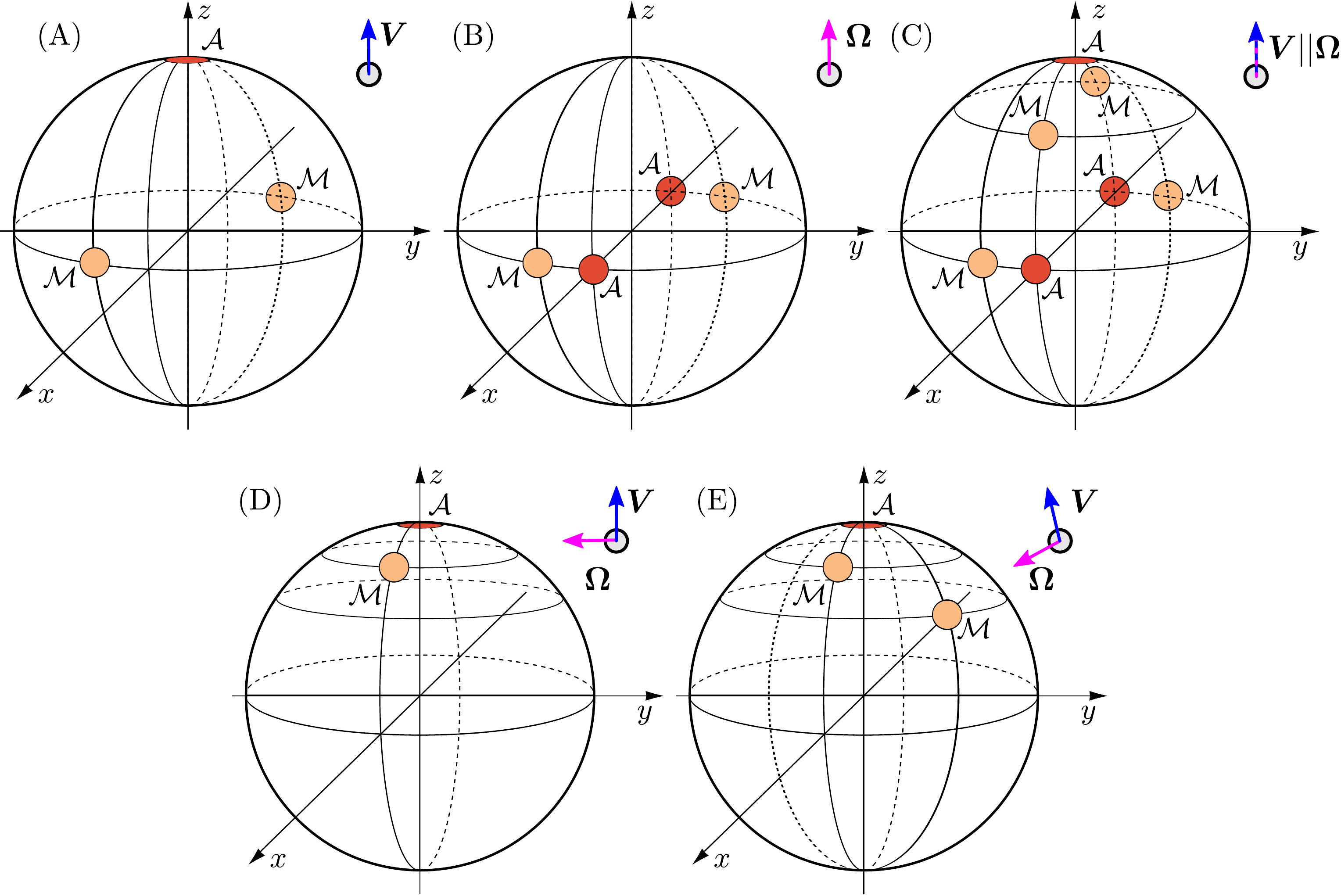}
\caption{Patch coverage (dark red active patches $\mathcal{A}$ and light orange  mobility
  patches $\mathcal{M}$) leading to the desired types of motion: (A) pure
  translation in the $z$-direction, (B) pure rotation about the $z$-axis, (C)
  collinear (independently controlled) translation and rotation, (D) translation perpendicular to the
  rotation axis, (E) a coverage with a single active post leading to
  translational and rotational motion such that $\bs{\Omega}\cdot\bs{V}\neq
  0$.
}
\label{patch}
\end{figure*}

The patch model can successfully reproduce the results for
continuous coverage given above. Simple examples of the patterns above are
sketched in Fig.~\ref{patch}. In Fig.~\ref{patch}(A) pure translation is
achieved by using a single active patch at the pole of the sphere and pairs of
identical mobility patches placed symmetrically so that the rotational effect
would be cancelled. In Fig.~\ref{patch}(B) using two activity patches and two
mobility patches lying on the circumference of a great circle with  a 2-fold
rotational symmetry leads to purely rotational motion.  Translation in the
direction of the rotation axis is achieved by adding a third activity patch to
the situation above at the pole, assuming the great circle to be the equator,
as in Fig.~\ref{patch}(C). The patches on the equator would suffice
    to produce both translational and rotational motion. An extra pair of
    patches in the Figure close to the pole allow to independently control
    the translational velocity.  A simple pair of patches, as introduced in the beginning of this section, and depicted in 
Fig.~\ref{patch}(D), leads to the linear and angular velocity in the body frame being orthogonal.

\begin{figure*}
 \centering
\includegraphics[width=0.55\textwidth]{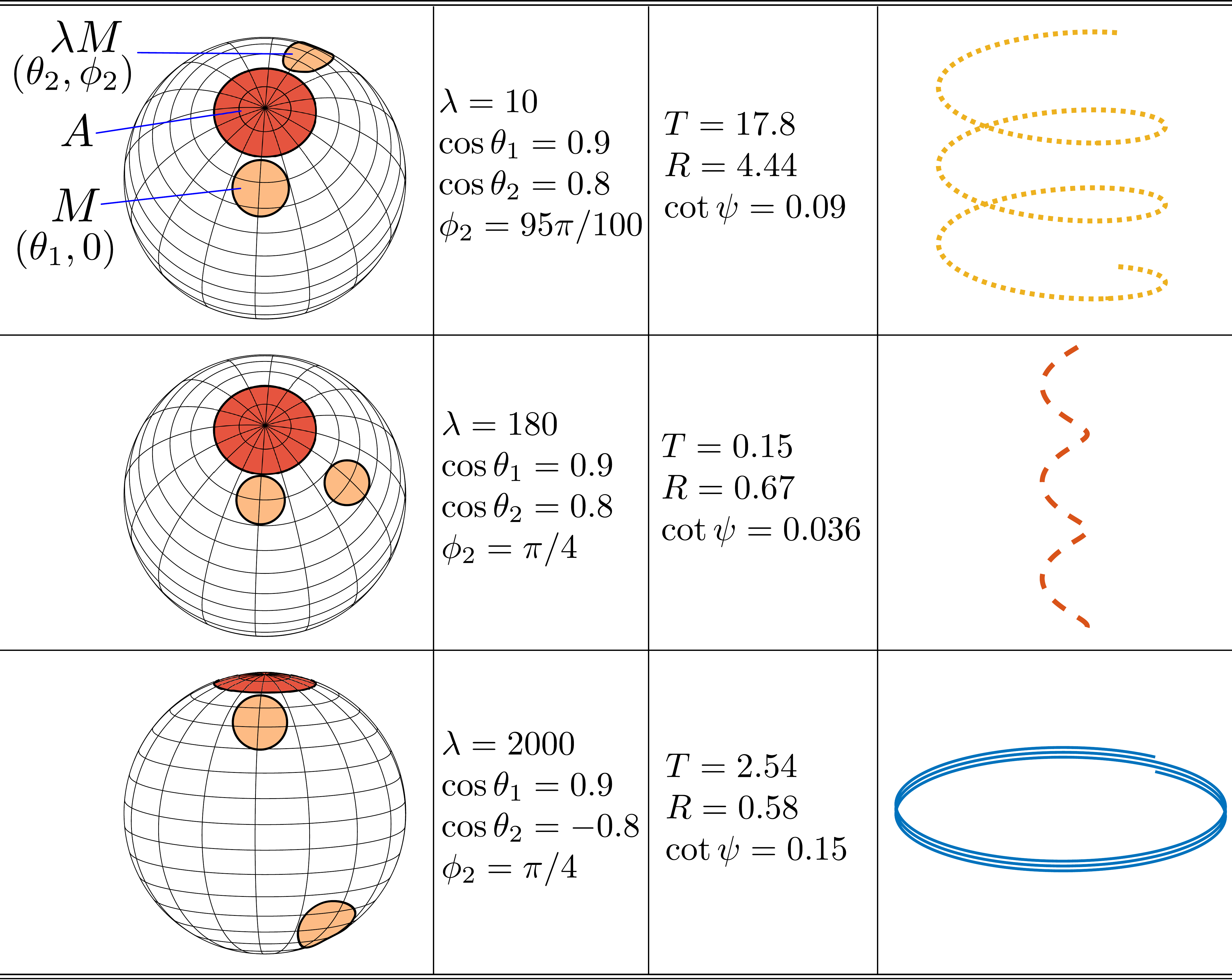}\includegraphics[width=0.45\textwidth]{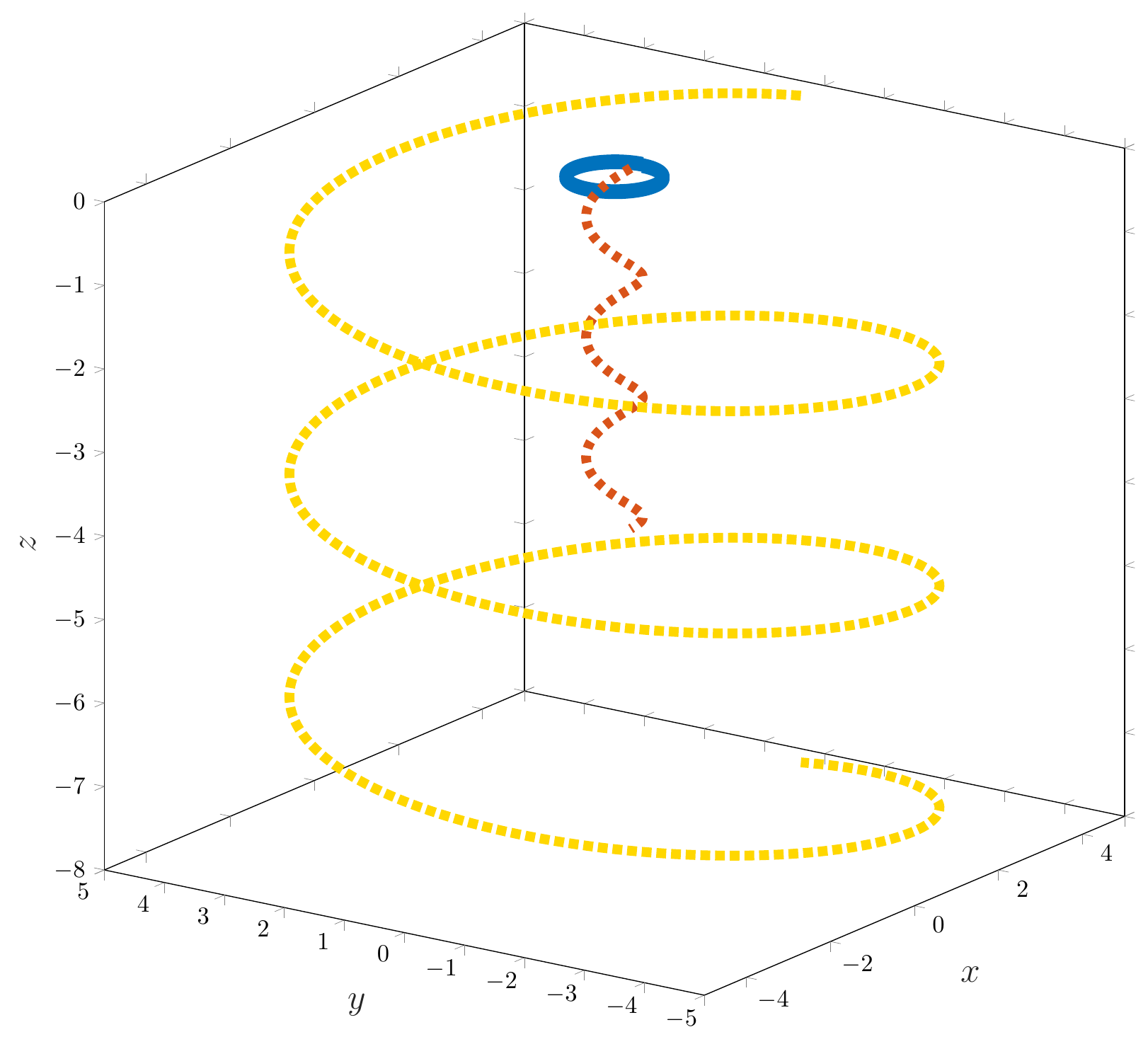}
\caption{Illustration of case (E) from Fig.~\ref{patch}: a sphere
  with one active patch $A$ and two patches of different mobilities, $M$ at
  $(\theta,\phi)=(\theta_1,0)$ and $\lambda M$ at $(\theta_2,\phi_2)$. Left:
  three cases considered with the parameters used for calculations. The
  resulting period of rotations ($T=2\pi/\Omega$), radius of the helix $R$ and
  its slope $\cot\psi$ are calculated in dimensionless units, with all lengths scaled by the particle radius $a$. Right: the trajectories with three full periods of rotation are plotted for each case
  with shapes and colours matching the examples on the left. }\label{trajectory}
\end{figure*}

General translations and rotations can be induced using many patches.  An exemplary coverage inducing rotation at an angle $\psi$ to translation is presented in Fig.~\ref{patch}(E) which we now detail.   The sphere is covered with a single patch of activity $A$ at the pole ($\theta=0$) and two patches of mobility, one of mobility $M$ at $(\theta,\phi)=(\theta_1,0)$ and a second one of mobility $\lambda M$ at $(\theta,\phi)=(\theta_2,\phi_2)$. The patches induce respective surface velocities $V_1 = MA G(\theta_1) /2D$ and $V_2 = \lambda MA G(\theta_2)/2D$ and angular velocities $\Omega_{1,2} = 3V_{1,2}/2a$.  The total translational and angular velocity due to the two pairs are thus
\begin{align}
\bs{V} &= \frac{MA}{2D}\begin{pmatrix}
G(\theta_1) \cos\theta_1 + \lambda G(\theta_2) \cos\theta_2\cos\phi_2 \\
\lambda G(\theta_2)\cos\theta_2\sin\phi_2 \\
G(\theta_1)\sin\theta_1+\lambda G(\theta_2)\sin\theta_2
\end{pmatrix}, \\
\bs{\Omega} &= \frac{3MA}{4aD} \begin{pmatrix}
\lambda G(\theta_2)\sin\phi_2 \\
G(\theta_1)+\lambda G(\theta_2)\cos\phi_2 \\
0
\end{pmatrix}.
\end{align}
The angle between the two is then determined as $\cos\psi =\frac{
  \bs{V}\cdot\bs{\Omega}}{|\bs{V}||\bs{\Omega}|}$. While the exact expression
is rather lengthy, indeed various angles are possible to engineer provided
that the system has enough asymmetry. For example, putting both mobility
patches at the same meridian ($\phi_2=0$) or putting them at the same circle
of longitude ($\theta_1=\theta_2$) leads to $\bs{V}$ and $\bs{\Omega}$ being
perpendicular. Interestingly, the speed of the particle is controlled by
  the factor $MA/D$, whereas the pitch angle of the resulting helix does not depend
  on these parametres. This suggests a possible way of tuning the particle
  trajectory and speed independently. The value of $\psi$ is sensitive to the choice of relative position and relative strength of the patches which in effect modifies the characteristics of the helical trajectory of the particle.  Examples of this coverage are illustrated in Fig.~\ref{trajectory}. We choose three specific patterns and demonstrate that, upon choosing the parameters suitably,  diverse trajectories can be produced. Results are presented in dimensionless variables. The scale for velocity is set by $V_0=MA/2D$, the angular velocity is scaled by $V_0/a$, with the particle size $a$ being the natural length scale in the system. Helices are characterised by their radius $R$ and slope $\cot\psi$, as derived in Appendix \ref{5kinematic}.
  
A more complicated arrangement of patches requires summations over all pairs,
yet modifying only the final angle $\psi$ which remains the only parameter that controls the geometry of swimming trajectories. The patch model provides easy qualitative arguments for the resultant motion given the position of functional sites. However, when the patches are extended into finite spots, the qualitative structure survives. Thus the arguments can be used to predict the motion created by a continuous spatial coverage, which would only change the quantitative characteristics of the motion.

\section{Conclusions}\label{6conclusions}

In this paper we have obtained two main results.  Firstly, we have developed
general formulae for the three-dimensional translational and rotational
velocity of spherical particles with  given arbitrary coverage in activity and
mobility. This allows to predict the resulting trajectories, which in general
are helical, although in the presence of thermal fluctuations the
  expected trajectory would be deviated by Brownian reorientation. These results are relevant to experimental observations of natural swimmers. Indeed, there are many examples of helical trajectories resulting from the wiggling motion due to the off-axis flagellar pushing \cite{Hyon2012}, or bacteria that inherently swim in helical trajectories: {\it Spiroplasma} \cite{Yang2009}, {\it Leptospira} \cite{Berg1979} or {\it Helicobacter pylori} \cite{Constantino2016}, to name a few. Although their helical motion originates from the anisotropic body shape, it is desirable to assess the effect of curvilinear paths on their dynamical interaction in bacterial suspensions. The route opened by our findings enables designing, and possibly manufacturing,  artificial swimmers that follow an arbitrary linear, circular, or helical path. The geometry is controlled by the relative magnitude and the angle between the linear and angular velocities, which can be calculated using our model. Pioneering experimental proof of this concept \cite{Archer2015} offers exciting possibilities of further practical applications of our results. 

{In this paper,  we neglected throughout  the effect of rotational Brownian motion which would lead to random reorientation of the translating and rotating phoretic particle.
Our model would still be appropriate to represent the trajectories at times short compared to the rotational diffusion time scale.}

For a continuous coverage, the task of evaluating the resulting velocities
might require the use of a computer to count in numerous terms in the
spherical harmonics expansion and evaluate integral expressions in the general
formulae. Moreover, it is not always clear at a first glance what
 the final motion would be given an arbitrary pattern of activity and mobility. To aid the intuition at this end, we have developed the patch model, which is the second important finding of the paper.

The proposed patch model had broad applications. We would like to stress that
increasing the sizes of patches would not alter the qualitative findings
of the paper but only modify the quantitative characteristics (provided that
patches do not overlap). A 'patch' need not imply only a small region of
non-zero mobility but  may be regarded as a small region of increased or
decreased mobility as compared to the surrounding surface, which can have a background value, i.e. a uniform surface mobility. This may be achieved by modifying surface charges on the particle, akin to systems in which electrostatic imbalance may lead to directed motion \cite{Yan2016}. The changing material properties across the surface can thus generate a distribution of activity and mobility on the surface that would be topologically equivalent to the one developed with the patch model. Our model could thus be used to help program the design of such particles. 

\section*{Conflicts of interest}
There are no conflicts to declare.

\section*{Acknowledgements}
This work has been funded in part  by the Ministry of Science and Higher Education of Poland via a Mobility Plus Fellowship  (M.L.),  the Foundation for Polish Science within the START programme (M.L.), the Isaac Newton Trust Cambridge (S.Y.R. and E.L.). This project has also received funding from the European Research Council (ERC) under the European Union's Horizon 2020 research and innovation programme  (grant agreement 682754 to E.L.).
\appendix
\section{Kinematics of a phoretic sphere}\label{5kinematic}

Consider a body-fixed frame $\{\bs{e}_x,\bs{e}_y,\bs{e}_z\}$. The phoretic effects result in the sphere having a fixed velocity $\bs{V}$ and angular velocity $\bs{\Omega}$ in this frame. Our aim is to find the lab frame motion of the particle. This can be done by evolving the body frame unit vectors in the inertial frame according to
\begin{align}
\td{\bs{e}_i}{t} &= \bs{\Omega}\times\bs{e}_i, \qquad i=x,y,z \\
\td{\bs{R}}{t} &= \sum_i V_i \bs{e}_i,  
\label{labtranslate}
\end{align}
where the unit vectors $\bs{e}_i$ and the centre of mass position $\bs{R}$ are expressed in the lab frame. The first set of equations can be conveniently written in a matrix form (with summation convention used henceforth) as
\begin{equation}
\td{\bs{e}_i}{t} = \mathsf{A}_{ij} \bs{e}_j,
\end{equation}
where 
\begin{equation}
\bs{\mathsf{A}} = \begin{pmatrix}
0 & \Omega_z & -\Omega_y \\
-\Omega_z & 0 & \Omega_x \\
\Omega_y & -\Omega_x & 0
\end{pmatrix},
\end{equation}
and the resulting linear system is conveniently solved using the matrix exponential $\bs{\mathsf{M}}(t)=\exp(\bs{\mathsf{A}}t)$ to give
\begin{equation}
\bs{e}_i(t) = \mathsf{M}_{ij}(t) \bs{e}_j(0).
\end{equation}
Assuming the body fixed frame to coincide with the lab frame at $t=0$, we find that the time evolution of the basis vectors $\bs{e}_i(t)$ is given by 
\begin{align}
\bs{e}_x(t) &= \frac{1}{\Omega^2}\begin{pmatrix}
\Omega_x^2 + (\Omega_y^2+\Omega_z^2)\cos\Omega t \\
\Omega_x\Omega_y(1-\cos\Omega t) + \Omega_z \Omega \sin\Omega t \\
\Omega_x\Omega_z(1-\cos\Omega t) - \Omega_y \Omega \sin\Omega t 
\end{pmatrix}, \\
\bs{e}_y(t) &= \frac{1}{\Omega^2}\begin{pmatrix}
\Omega_x\Omega_y(1-\cos\Omega t) - \Omega_z \Omega \sin\Omega t \\
\Omega_y^2 + (\Omega_x^2+\Omega_z^2)\cos\Omega t \\
\Omega_y\Omega_z(1-\cos\Omega t) + \Omega_x \Omega \sin\Omega t 
\end{pmatrix}, \\
\bs{e}_z(t) &= \frac{1}{\Omega^2}\begin{pmatrix}
\Omega_x\Omega_z(1-\cos\Omega t) + \Omega_y \Omega \sin\Omega t \\
 \Omega_y\Omega_z(1-\cos\Omega t) - \Omega_x \Omega \sin\Omega t \\
\Omega_z^2 + (\Omega_x^2+\Omega_y^2)\cos\Omega t
\end{pmatrix}.
\end{align}
Thus Eq.~\eqref{labtranslate} can be integrated to yield the position of the centre of mass in the laboratory frame:
\begin{align}
 \bs{R}(t) = \bs{P}t + \bs{Q}\cos\Omega t + \bs{S} \sin\Omega t.
\end{align}
with constant vectors $\bs{P},\bs{Q},\bs{S}$ determined by the initial values
of velocities $\bs{V}_0$ and $\bs{\Omega}_0$ in the lab frame. Henceforth, we
drop the index for brevity assuming that the velocity vectors are taken at
$t=0$ in the laboratory frame. If the the angle between them is $\psi$, we
have $\bs{V}\cdot\bs{\Omega} = V \Omega \cos\psi$ and $\bs{V}\times\bs{\Omega}
= V \Omega \sin\psi  \bs{\hat{U}}$, where $V=|\bs{V}|$, $\Omega=|\bs{\Omega}|$
and $\bs{\hat{U}}$ is a unit vector (which we denote by hats) perpendicular to
both $\bs{V}$ and $\bs{\Omega}$. Introducing the unit vector
$\bs{\hat{W}}=\bs{\hat{\Omega}}\times\bs{\hat{U}}$, we see that
$\{\bs{\hat{\Omega}},\bs{\hat{U}},\bs{\hat{W}}\}$ form an orthonormal basis. 
With this basis, we find
\begin{align}
\bs{P} &= V  \cos\psi \bs{\hat{\Omega}}, \\
\bs{Q} &= \rho\sin\psi \bs{\hat{U}}, \\
\bs{S} &= \rho \sin\psi \bs{\hat{W}}= \rho\left(\bs{\hat{V}} - \bs{\hat{\Omega}} \cos\psi \right),
\end{align}
where $\rho=V/\Omega$. The trajectory becomes thus
\begin{equation}
\bs{R}(t)=   V t \cos\psi \bs{\hat{\Omega}} + \rho \sin\psi \left(\bs{\hat{U}}\cos\Omega t + \bs{\hat{W}} \sin\Omega t \right) .
\end{equation}

The simplest case of motion is when $\bs{V}$ and $\bs{\Omega}$ are collinear, that is for $\psi=0$. The motion is then along a straight line determined by the direction of $\bs{V}$ being also the rotational axis. Another interesting case is when the translational and rotational velocities are perpendicular, $\bs{V}\cdot\bs{\Omega}=0$ and and $\psi=\pi/2$. Then $\bs{P}=0$ and the motion is determined by $\bs{Q} = \rho \bs{\hat{U}}$ and $\bs{S} = \rho \bs{\hat{V}}$, and we find the trajectory 
\begin{equation}
 \bs{R}(t) = \rho \left[\bs{\hat{V}} \sin\Omega t + \bs{\hat{U}} \cos\Omega t\right],
\end{equation}
which describes a circle of radius $\rho$ in the plane spanned by the direction of velocity and the direction perpendicular to both velocity vectors. In the most general case, the trajectory is helical with the axis along the direction of $\bs{\Omega}$, the helix radius $\rho\sin\psi$ and pitch $2\pi \rho \cos\psi$ (or slope $\cot\psi$).

\end{document}